\documentclass[aps,twocolumn,superscriptaddress,floatfix,longbibliography]{revtex4-2}
\usepackage{amsmath,amssymb,amsthm}
\usepackage{physics}
\usepackage{amsfonts}
\usepackage{mathrsfs}
\usepackage{graphicx}
\usepackage{tabularx}
\usepackage{enumerate}
\usepackage{dcolumn}
\usepackage{bm}
\usepackage{xcolor}
\usepackage[normalem]{ulem}
\usepackage[colorlinks,linkcolor=blue,citecolor=blue,urlcolor=blue]{hyperref}

\begin{document}
\title{Dissipation induced ergodic-nonergodic transition in finite-height mosaic Wannier-Stark lattices}







\author{Xiang-Ping Jiang}
\affiliation{Zhejiang Lab, Hangzhou 311121, China}

\author{Xuanpu Yang}
\affiliation{School of Physics, Nankai University, Tianjin 300071, China}
\thanks{The first two authors contributed equally to this work.}

\author{Yayun Hu}
\email{yyhu@zhejianglab.edu.cn}
\affiliation{Zhejiang Lab, Hangzhou 311121, China}

\author{Lei Pan}
\email{panlei@nankai.edu.cn}
\affiliation{School of Physics, Nankai University, Tianjin 300071, China}

\date{\today}

\begin{abstract}
Recent research has observed the occurrence of pseudo-mobility edge (ME) within a modulated mosaic model incorporating the Wannier-Stark potential. This pseudo-ME, which signifies the critical energy that distinguishes between ergodic and weakly ergodic, or weakly ergodic and nonergodic states, is a crucial concept in comprehending the transport and localization phenomena in  Wannier-Stark systems. Here we investigate the influence of dissipation on a finite-height mosaic Wannier-Stark lattice that features such pseudo-MEs by computing the steady state density matrix. Our findings indicate that particular dissipation can steer the system into specific states, regardless of its initial state, predominantly characterized by either ergodic or nonergodic states. This suggests that dissipation can be harnessed as a novel method for inducing transitions between these states and manipulating particle localization behaviors in disorder-free systems.
\end{abstract}

\maketitle

\section{Introduction}
The investigation into the localization of electronic Bloch waves, a phenomenon where electronic wave functions become exponentially confined due to disorder, was initially introduced by P. W. Anderson~\cite{anderson1958absence}. In a three-dimensional (3D) disordered system, a mobility edge (ME) emerges, which acts as a critical energy separating localized and extended states as a function of the disorder level~\cite{lee1985disordered,kramer1993localization,evers2008anderson}. The disorder-induced localization transition is typically confined to dimensions greater than two, by scaling theory~\cite{thouless1974electrons,abrahams1979scaling,hetenyi2021scaling}; however, it can also be observed in one-dimensional (1D) quasiperiodic systems. For example, the 1D Aubry-Andr{\'e}-Harper (AAH) model~\cite{harper1955single,aubry1980analyticity} demonstrates the existence of an energy-independent extended-localized transition. By further generalizing the AAH models, such as varying the onsite potential~\cite{sarma1988mobility,ganeshan2015nearest}, incorporating long-range hopping~\cite{biddle2010pre,deng2019one}, or introducing a quasiperiodic potential in mosaic lattices~\cite{wang2020one,zhou2023exact}, these modified models can support an energy-dependent ME in the energy spectrum~\cite{liu2018mobility,zhang2022lyapunov,liu2022anomalous,liu2020non,xia2022exact,qi2023multiple,gonccalves2023renormalization,gonccalves2023critical,wang2023two,jiang2024exact}.

In a recent study~\cite{dwiputra2022single}, it has been proposed that a disorder-free 1D mosaic lattice, incorporating the Wannier-Stark potential, may exhibit an exact ME. This phenomenon can be traced back to the renowned Wannier-Stark ladder. By introducing a static electric field into the lattice, the resulting potential is known to lead to exponential localization of the system's eigenstate. The Wannier-Stark ladder is recovered with sufficiently strong fields, and each energy level corresponds to a Wannier-Stark localized eigenstate~\cite{wannier1962dynamics,fukuyama1973tightly,emin1987existence,zeng2023wannier,hartmann2004dynamics,zhang2022engineering}. Despite the excitement surrounding this theory, recent work~\cite{longhi2023absence} by S. Longhi has raised some concerns. According to Longhi, Avila's global theory~\cite{avila2015global} cannot be applied in this context, and the definition of Lyapunov exponents becomes problematic for Wannier-Stark potentials that approach infinity. In the thermodynamic limit, where the system size approaches infinity, all states become localized, with only a few isolated extended states remaining. Therefore, strictly speaking, no true disorder-free ME exists. Nonetheless, within a mosaic potential possessing a finite height, the Wannier-Stark lattice can exhibit a pseudo-ME, indicating the critical energy distinguishing between ergodic and weakly ergodic, or weakly ergodic and nonergodic states. The coexistence of ergodic and nonergodic states in finite-sized mosaic Wannier-Stark lattices has been demonstrated experimentally~\cite{guo2021observation,gao2023coexistence} and investigated theoretically~\cite{mendez1988stark,voisin1988observation,mendez1993wannier,wei2022static,qi2023localization,wei2024coexistence}.

Non-Hermitian physics has experienced significant advancements, particularly in manipulating dissipation and quantum coherence in experimental settings~\cite{luschen2017signatures,xiao2020non,liang2022dynamic,gao2024experimental}. These progresses have triggered a growing interest in the study of dissipative open quantum systems~\cite{yamamoto2022universal,mao2023non,mao2024liouvillian,zheng2024exact,ekman2024liouvillian,qin2024occupation}. Dissipation is acknowledged as a potent force capable of altering fundamental properties of quantum systems, resulting in various phase transitions~\cite{prosen2008quantum,mebrahtu2012quantum,longhi2019topological,hamazaki2019non,xu2020topological,liuT2020non,shastri2020dissipation,nie2021dissipative,yamamoto2021collective,zeng2020topological1,zeng2020topological2,weidemann2022topological,wu2021non,li2023non,zhu2023topological,liuT2020non,gandhi2023topological,liu2023ergodicity,kawabata2023entanglement,li2024emergent,yu2024non,zhou2024entanglement1,zhou2024entanglement2,jing2024biorthogonal}. The influence of dissipation on localization and transport characteristics within disordered and quasiperiodic systems has also become a central focus of extensive research~\cite{gurvitz2000delocalization,yamilov2014position,huse2015localized,balasubrahmaniyam2020necklace,weidemann2021coexistence,yusipov2017localization,purkayastha2017nonequilibrium,vershinina2017control,yusipov2018quantum,vakulchyk2018signatures,balachandran2019energy,chiaracane2020quasiperiodic,lacerda2021dephasing,chiaracane2022dephasing,dwiputra2021environment,longhi2023anderson,longhi2024dephasing,liu2024dissipation}. Recent studies have demonstrated that dissipation can drive Anderson localization into a stable state that retains its localized properties without being destroyed. Moreover, for 1D quasiperiodic systems with MEs, dissipation can drive the system into specific states, which may be extended or localized regardless of the initial state. This suggests that dissipation can induce transitions between extended and localized states. However, to date, studies on the effects of dissipation on the localization and transport behaviors of disorder-free systems with pseudo-MEs have not been explored fully.

\begin{figure*}[t]
	\centering
	\includegraphics[width=0.32\textwidth]{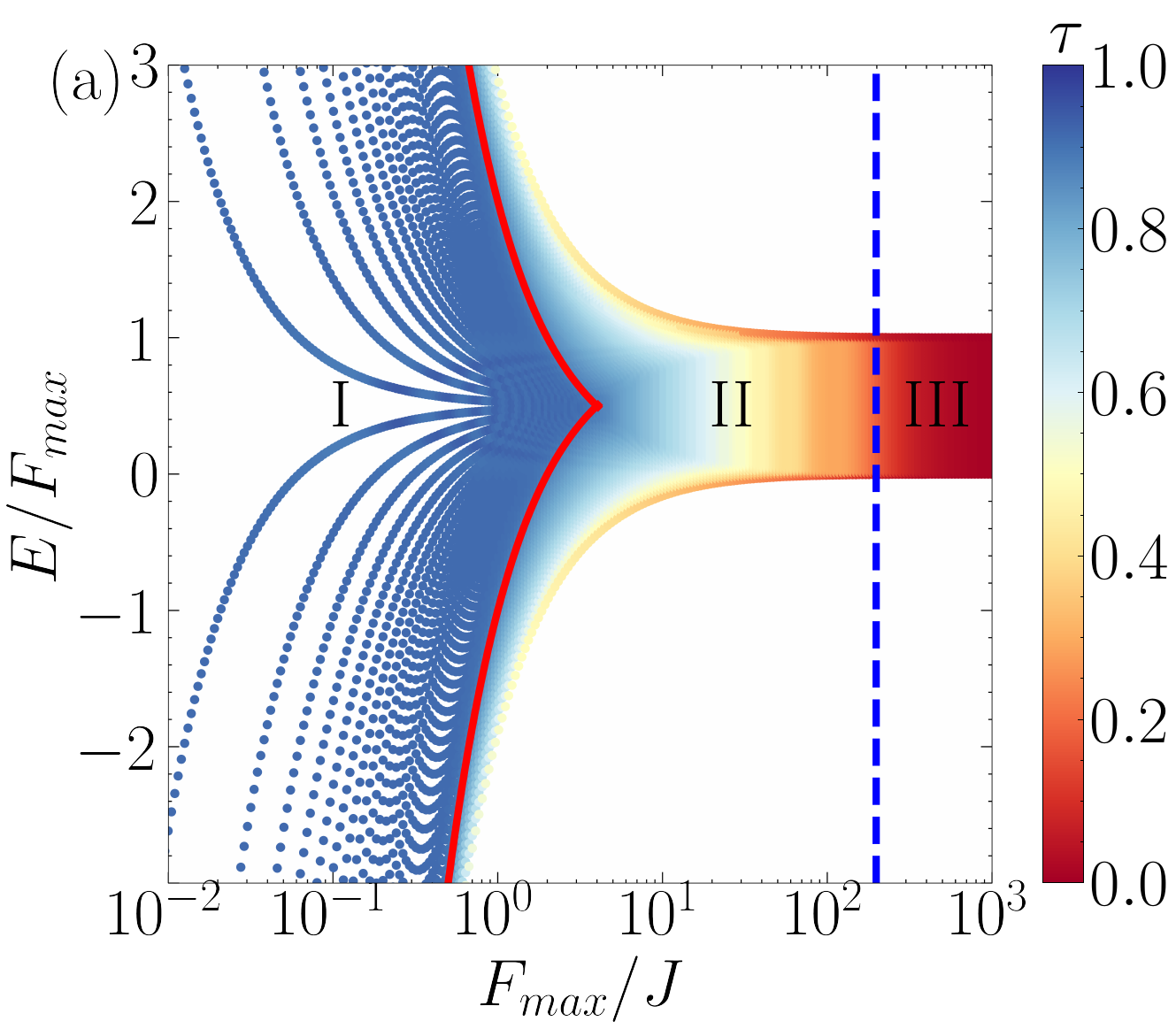}
	\includegraphics[width=0.32\textwidth]{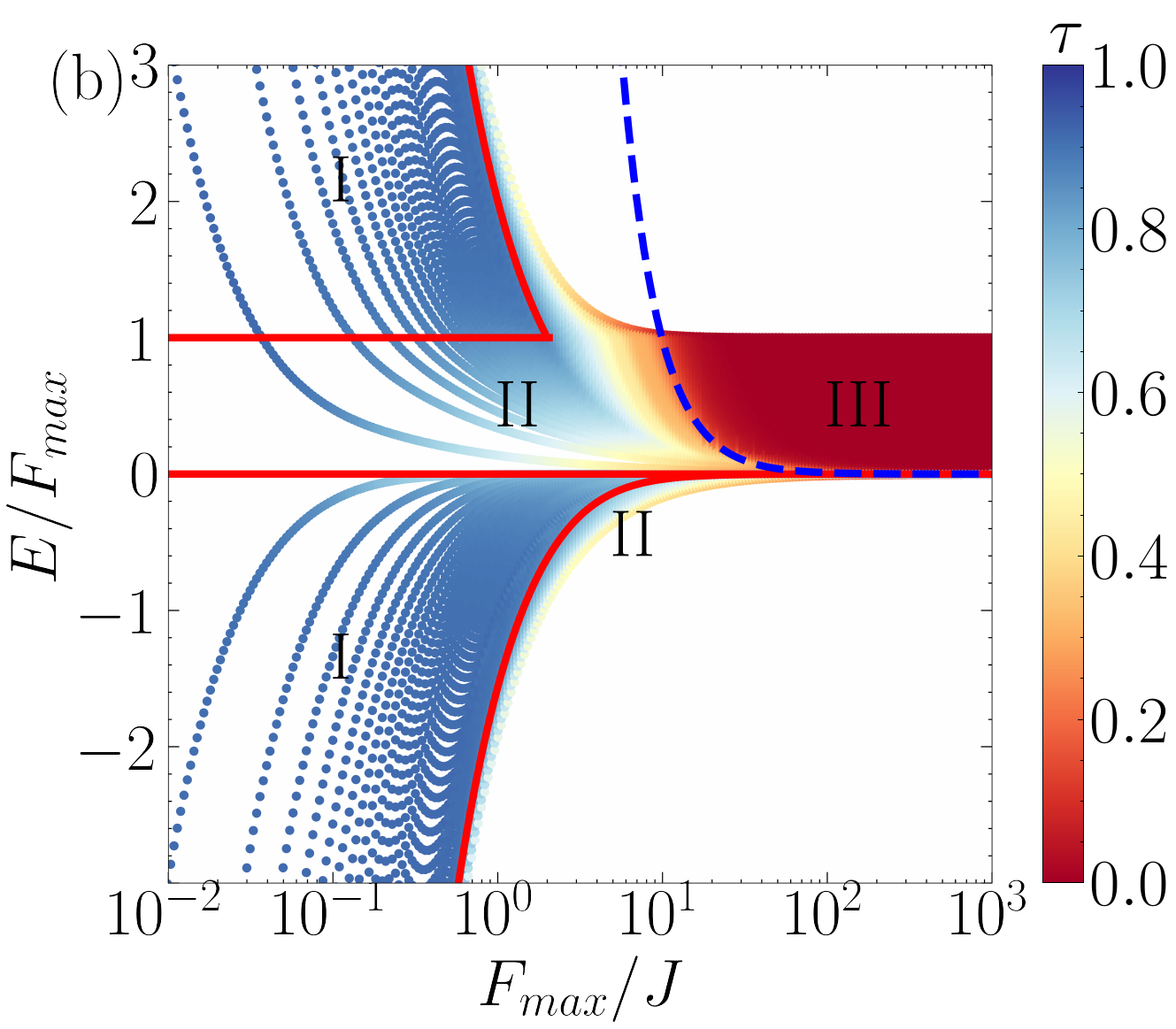}
    \includegraphics[width=0.34\textwidth]{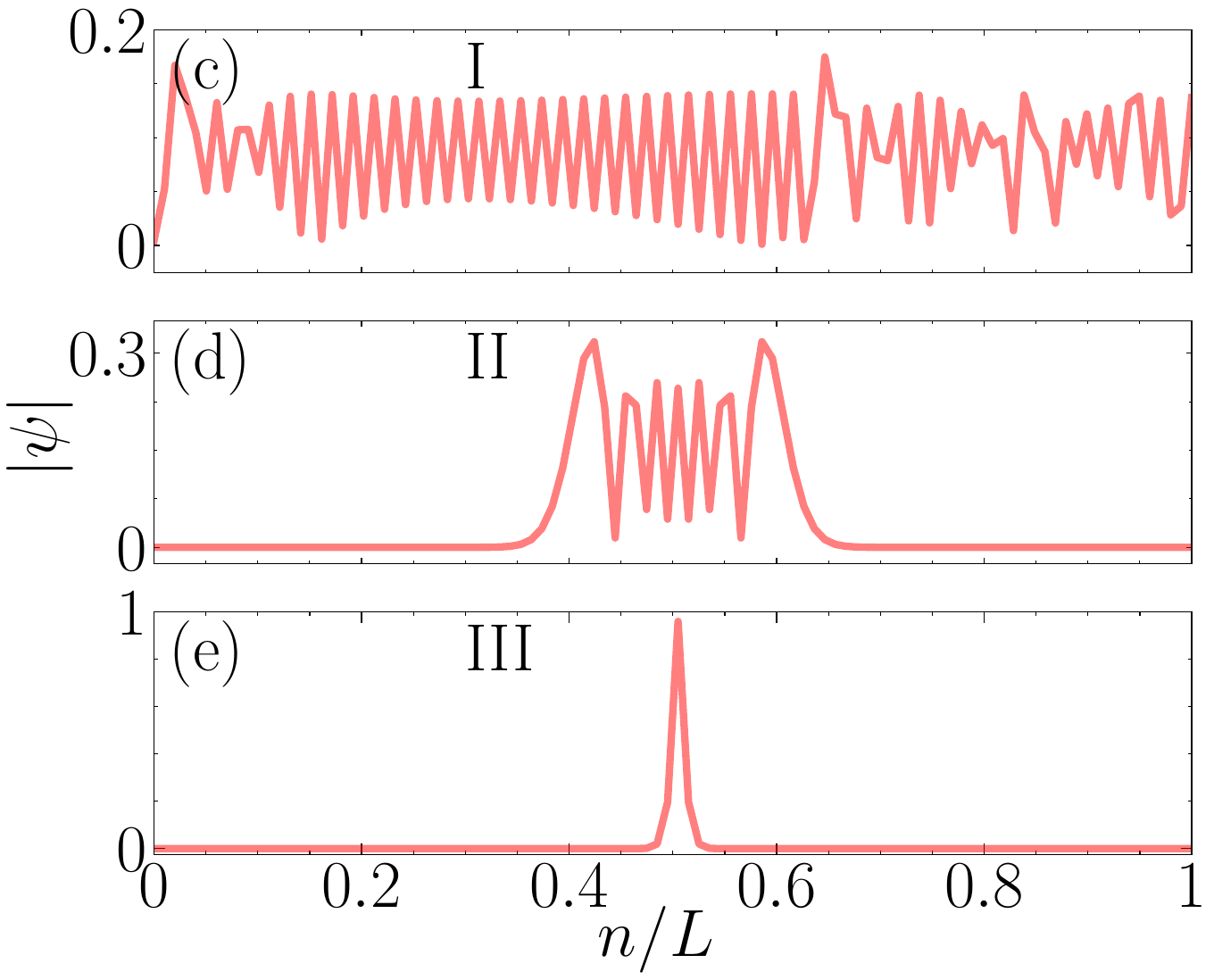}
	\caption{Localization and the pseudo-mobility edge (pseudo-ME) of the finite-height mosaic Wannier-Stark lattice. (a), (b) are the entire phase diagram of the Hamiltonian (\ref{Hamiltonian}) with the $ \kappa=1 $ and $ \kappa=2 $ case, respectively. Fractal dimension (FD) $ \tau $ of different eigenstates and the corresponding energy $E/F_{max}$ as a function of $ F_{max}/J $ under open boundary conditions. The red solid lines represent the critical energies between ergodic and weak ergodic states, while the blue dashed line marks the critical energy separating from weak ergodic to nonergodic states. These pseudo-MEs separate the energy spectrum into three regions: $ \tau \simeq 1 $ for the ergodic states in the region I, $ 0<\tau< 1 $ for the weakly ergodic states in the region II, and $ \tau \simeq 0 $ for the nonergodic states in the region III. (c-e) The spatial distribution of typical wave functions for ergodic states, weakly ergodic states, and strongly Wannier-Stark-localized (nonergodic) states corresponds to the center energy for the $\kappa=1$ case with $F_{max}/J=0.1,20,500$, respectively. The chain length is $L=100$.}
	\label{fig1}
\end{figure*}

In this work, we explore the impact of dissipation on a 1D finite-height mosaic Wannier-Stark lattice incorporating pseudo-MEs. Our findings reveal that despite the initial state, dissipation can guide the system towards specific steady states, which can be either ergodic, weakly ergodic, or nonergodic. This observation implies that dissipation induces transitions between ergodic states and weakly ergodic or nonergodic states, and between weakly ergodic states and ergodic or nonergodic states in disorder-free systems with pseudo-MEs. Consequently, the combination of dissipation and pseudo-MEs yields novel and intricate physical phenomena, offering an innovative approach to manipulating the ergodic-nonergodic transition and transport properties of particles in disorder-free systems.

The organization of this paper is as follows: In Sec. \ref{sec: models}, we delineate the finite-height mosaic Wannier-Stark lattice equipped with pseudo-MEs. In Sec. \ref{sec: Lindblad}, we consider a dissipative system following the Lindblad master equation with the corresponding jump operator. The influence of dissipation on our disorder-free systems is explored in Sec. \ref{sec: results}. Finally, a comprehensive summary of the findings is provided in Sec. \ref{sec: conclusion}.

\section{The model Hamiltonian}\label{sec: models}
We consider a 1D tight-bound lattice with the mosaic Wannier-Stark potential under open boundary conditions, as depicted by the Hamiltonian~\cite{dwiputra2022single,longhi2023absence,wei2024coexistence}:
\begin{eqnarray}\label{Hamiltonian}
H=J\sum_{j=1}\left(c_{j}^{\dagger}c_{j+1}+c_{j+1}^{\dagger}c_{j}\right)+\sum_{j=1}V_{j}c_{j}^{\dagger}c_{j},
\end{eqnarray}
where $c^{\dagger}_j$ and $c_j$ represent the fermionic creation and annihilation operators at site $j$, and $J$ is the nearest neighbor hopping strength. Here, $V_j$ is the onsite potential at site $j$, which takes the following form:
\begin{eqnarray}\label{eq:Disord} 
V_j=\left\{ 
\begin{array}{cl}
	F j , & j=\kappa n  ,\\
	0,& \mathrm{otherwise},\\
\end{array} 
\right.
\end{eqnarray}
where $F$ and $\kappa$ denote the electronic field and the mosaic periodic parameter, respectively. The potential's non-uniformity prompts us to define a supercell encompassing every $\kappa$ site. If the supercell number of the system is denoted as $N$, i.e., $n = 0,1,2,..., N-1$, the lattice length will be $L=\kappa N$. Our model's maximum potential value is $F_{max}=\kappa F(N-1)$. In the present work, we consider the finite-height potential, i.e., $F_{max}$ is finite and independent of the size of the system.  Without loss of generality, we set $J=1$ as the energy unit and choose $\kappa=1$ and $\kappa=2$, corresponding to the linear and mosaic potential, respectively. For $\kappa=1$, the Hamiltonian (\ref{Hamiltonian}) simplifies to the well-known Wannier-Stark model, whose eigenstates are all localized states for any finite potential in the thermodynamic limit. For $\kappa=2$, the presence of pseudo-MEs has been recently reported. It is noteworthy that the finite-size mosaic Wannier-Stark models have garnered considerable attention in the realm of theoretical research in recent times. Moreover, these models have proven effective in various experimental setups, including superconducting qubits and photonic devices.

To more accurately describe the properties of eigenstates, we further investigate the wave function's fractal dimension (FD), which is associated with the inverse participation ratio (IPR)~\cite{qi2023multiple,jiang2021mobility,wang2023two}. We consider the eigenquation $H|\psi_n\rangle=E|\psi_n\rangle$, where the $n$th eigenstate is $|\psi_n\rangle=\sum_{j}\psi_{n,j}c_{j}^{\dagger}|0\rangle$ and $\psi$ is the eigenstate amplitude. Thus the corresponding ${\rm{IPR}} =\sum_j|\psi_j|^4$, and the FD is defined as
\begin{equation}\label{fd}
\begin{aligned}
\tau=-\frac{\ln \text{(IPR)}}{\ln L},
\end{aligned}
\end{equation}
Within the thermodynamic limit, the FD $\tau$ assumes distinct values corresponding to different states: $\tau=1$ for the extended (ergodic) states, $\tau=0$ for the localized (nonergodic) states, and $0<\tau<1$ for the critical states, which are characterized by multifractal properties typically observed in disordered or quasiperiodic systems. Notably, for our Hamiltonian (\ref{Hamiltonian}) without the disorder, we show that the entire phase diagram can partition the energy spectra into three distinct regions: ergodic regions, weakly ergodic regions, and strongly Wannier-Stark localized (nonergodic) regions. We will present this intricate phase diagram in the subsequent discussion.

As shown in Refs.~\cite{gao2023coexistence,wei2024coexistence}, we can analytically obtain the pseudo-MEs for the finite-height mosaic Wannier-Stark models. The main numerical results are shown in Fig. \ref{fig1}. For the Hamiltonian (\ref{Hamiltonian}) with $\kappa=1$ case, the pseudo-MEs are
\begin{equation}\label{k1}
E/F_{max}=\left\{\begin{aligned}
&2/F_{max}, \\
&1-2/F_{max},\\
\end{aligned}\right.
\end{equation}
and $E/F_{max}=2(N-1)$, respectively. In the former case, as shown in Fig. \ref{fig1}(a), these two critical energies (red solid lines) separate the ergodic states (region I) from weakly ergodic states (region II), while in the latter case, the critical energy (blue dashed line) marks the boundary between weakly ergodic states and nonergodic states (region III). In Fig. \ref{fig1}(b), we consider the mosaic potential $\kappa=2$ and the pseudo-MEs are
\begin{equation}\label{k2}
E/F_{max}=\left\{\begin{aligned}
&0, \\
&2/F_{max}, \\
&\frac{1-\sqrt{1+16/F_{max}^2}}{2},\\
&1,\\
\end{aligned}\right.
\end{equation}
and $E/F_{max}=2(N-1)/F^{2}_{max}$, respectively. In the former case, these four critical energies (red solid lines) separate the ergodic states (region I) from weakly ergodic states (region II), while in the latter case, the critical energy (blue dashed line) marks the boundary between weakly ergodic states and nonergodic states (region III). To more accurately describe typical properties of eigenstates for ergodic states, weakly ergodic states, and nonergodic states, we further plot the distribution of the wave functions for $\kappa=1$ case with $F_{max}/J=0.1,20,500$ in Fig. \ref{fig1}(c-e). Consequently, we have completed the entire phase diagram of the finite-height mosaic Wannier-Stark system and determined the three regions: $ \tau \simeq 1 $ for the ergodic states (region I), $ 0<\tau < 1 $ for the weakly ergodic states (region II), and $ \tau \simeq 0 $ the nonergodic states (region III). 

\begin{figure*}[t]
	\centering
	\includegraphics[width=0.65\textwidth]{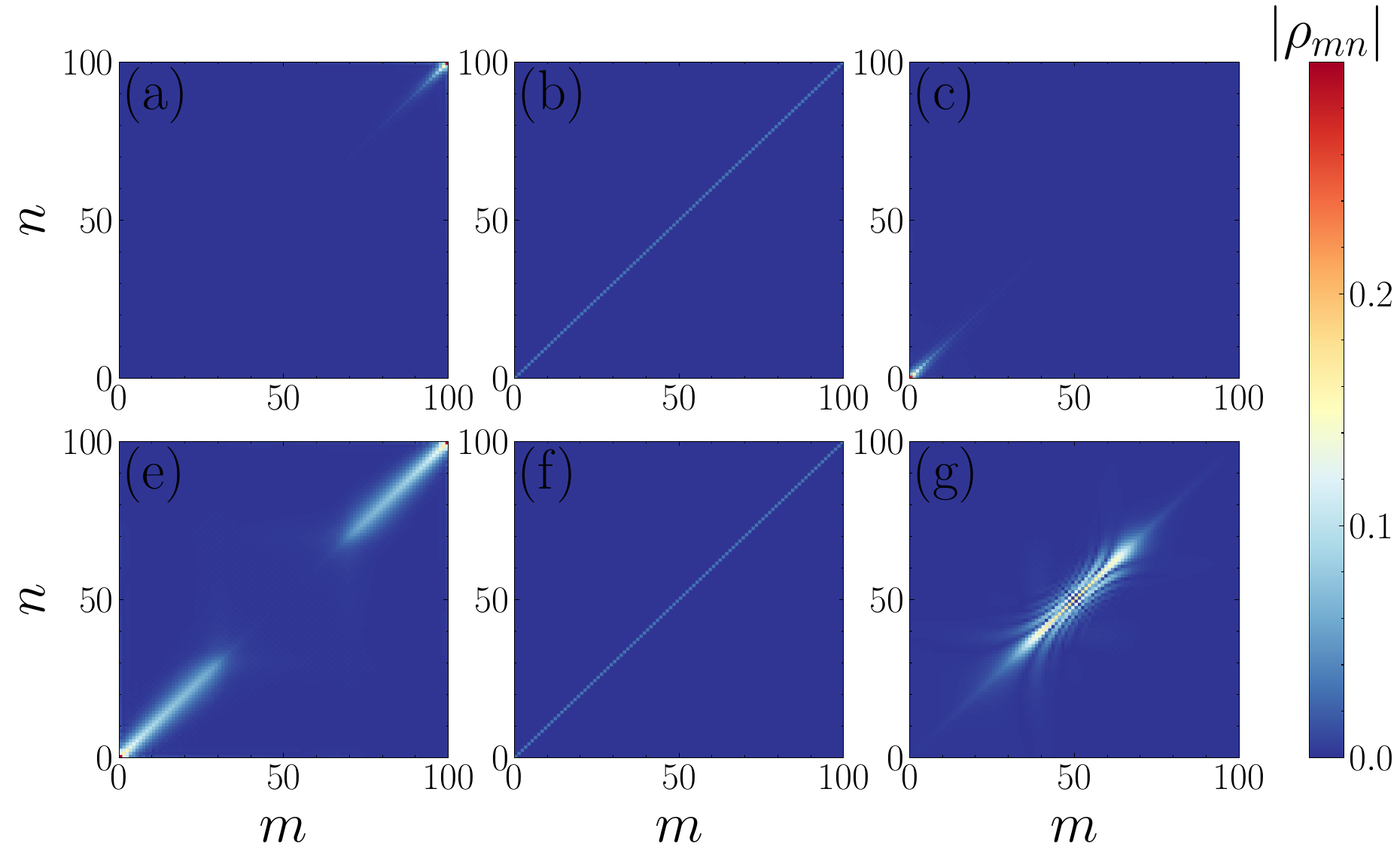}
    \includegraphics[width=0.31\textwidth]{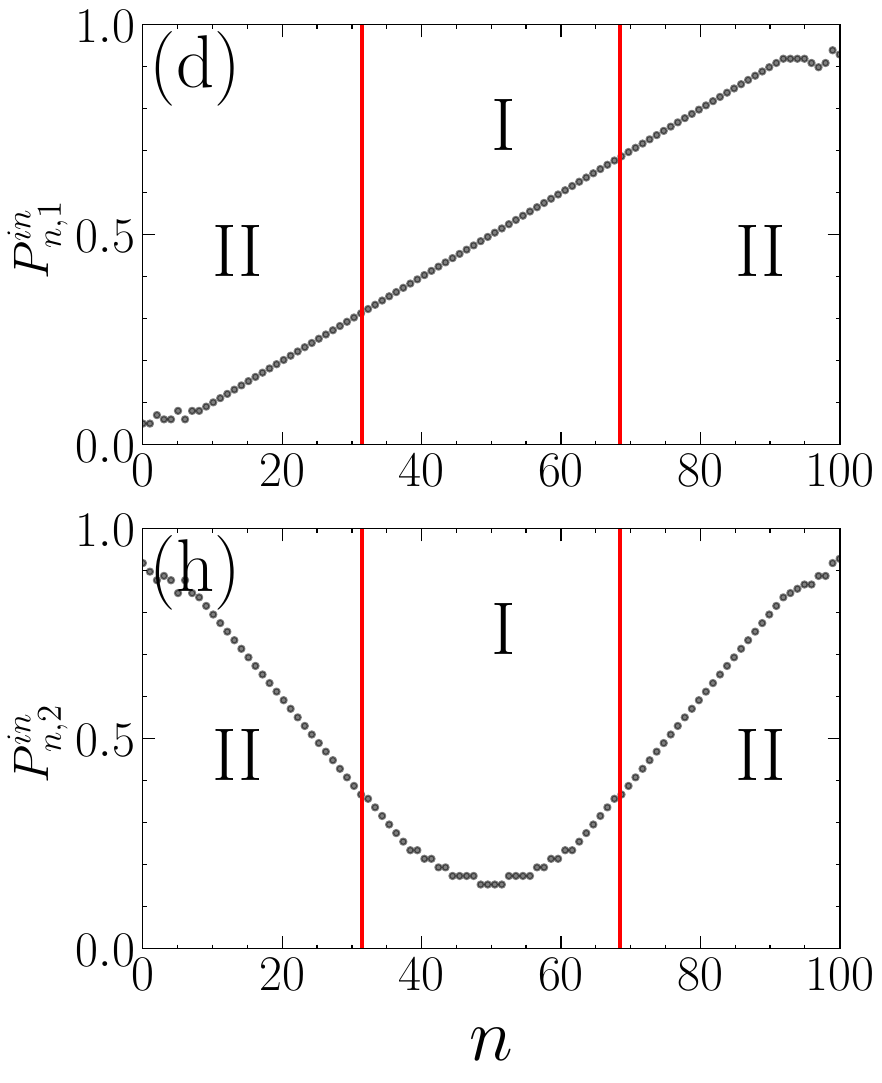}
	\caption{The density matrix distributions for steady states with the different dissipation phases (a) $\theta=0$, (b) $\theta=\pi/2$, and (c) $\theta=\pi$ in the eigenstates basis of the Hamiltonian (\ref{Hamiltonian}) with $\kappa=1$ and $l=1$. (d) The fraction of in-phase pairs $P_{n,1}^{\text{in}}$ for different eigenstate. The red solid lines mark the pseudo-MEs, separating eigenstates into weak ergodic (II), ergodic (I), and weak ergodic (II) regions as eigenvalues increase. The density matrix distributions for steady states with the different dissipation phases (e) $\theta=0$, (f) $\theta=\pi/2$, and (g) $\theta=\pi$ in the eigenstates basis of the Hamiltonian (\ref{Hamiltonian})  with $\kappa=1$ and $l=2$. (h) The fraction of in-phase pairs $P_{n,2}^{\text{in}}$ for different eigenstates. Here we take $L=100$, $\Gamma=1$, and $F_{max}/J=2$.}
	\label{fig2}
\end{figure*}

\begin{figure}[b!]
	\centering
	\includegraphics[width=0.48\textwidth]{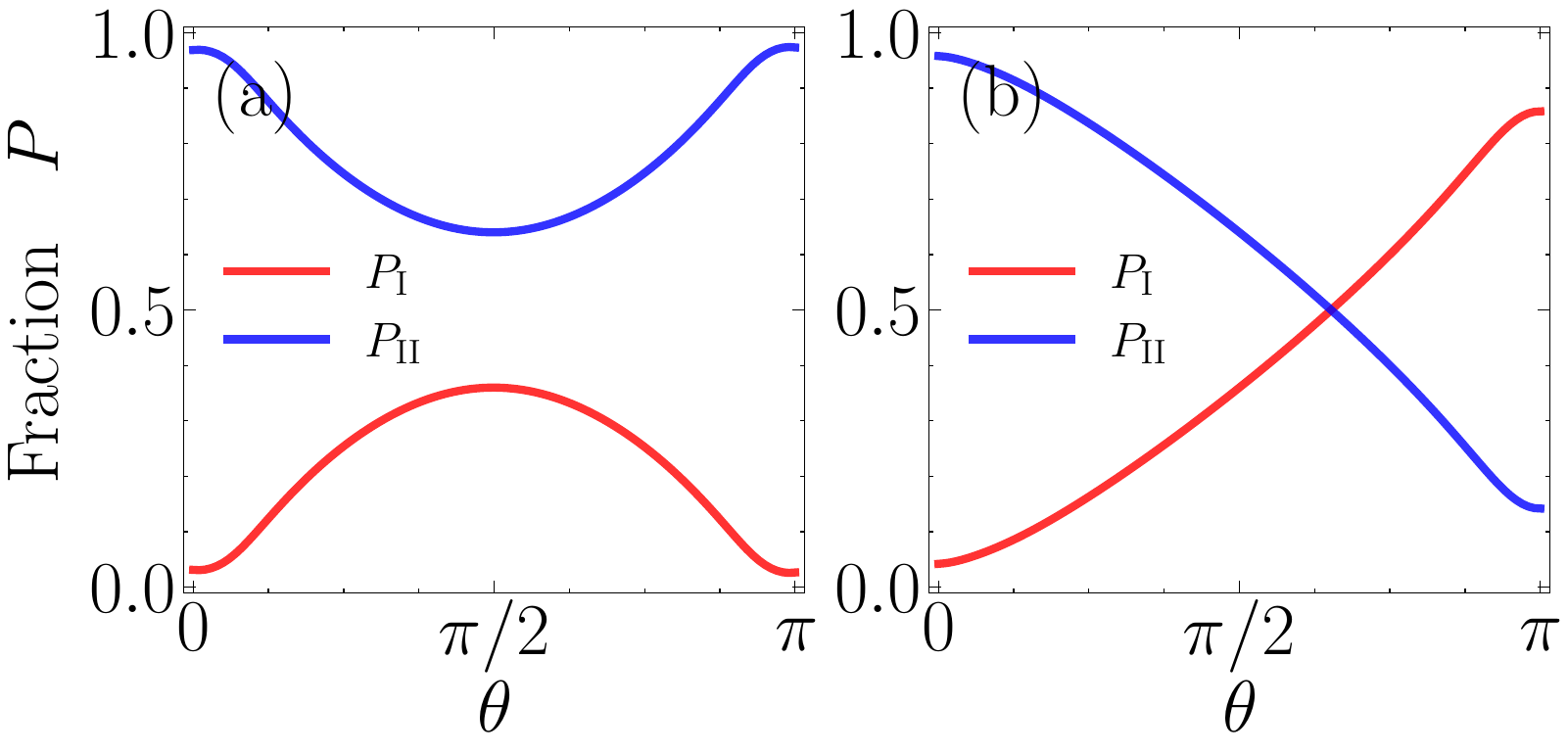}
	\caption{Fractions of ergodic eigenstates $P_{\rm{I}}$, weakly ergodic eigenstates $P_{\rm{I}}$ and weakly ergodic eigenstates $P_{\rm{II}}$ in corresponding steady states as the function of dissipatiion phase $\theta$ for the Hamiltonian (\ref{Hamiltonian}) with $\kappa=1$. (a) and (b) show the results for $l=1$ and $l=2$ case, respectively.}
	\label{fig3}
\end{figure}

\section{The Lindblad master equation and the jump operator}\label{sec: Lindblad}
In this section, we consider a general dissipative system whose density matrix $\rho(t)$ follows the following Lindblad master equation~\cite{lindblad1976generators}, 
\begin{align}
	\frac{d\rho(t)}{dt}  \equiv \mathscr{L} [\rho(t)]= -i\left[H, \rho(t)\right] +\mathcal{D}[ \rho(t)],
 \label{LindbladEq}
\end{align}
where $\mathscr{L}$ is the Liouvillian superoperator or Lindbladian. The associated dissipator is expressed as
\begin{eqnarray}\label{Dj}
	\mathcal{D}[\rho(t)] =  \sum_j\Gamma_{j}\left(\ell_{j}\rho \ell_{j}^{\dagger} -\frac{1}{2}\{ \ell_{j}^{\dagger} \ell_{j} , \rho \}\right),
\end{eqnarray}
which consists of quantum jump operators $\ell_{j}$ with the dissipation strengths $\Gamma_{j}$. If the Liouvillian superoperator $\mathscr{L}$ is time-independent, the formal solution is expressed as $\rho(t)=e^{\mathscr{L}t}\rho(0)$. Generally, the system will relax to a steady state which is defined by $\rho_{ss}=\lim_{t\rightarrow\infty}\rho(t)$, corresponding to the zero-eigenvalue eigenstate of $\mathscr{L}$, namely $\mathscr{L}[\rho_{ss}]=0$. 

Based on the Choi-Jamiolkowski isomorphism~\cite{jamiolkowski1972linear,choi1975completely}, we can rewrite the master equation \eqref{LindbladEq} as the form $\frac{d}{dt}{|\rho(t)\rangle}= \mathscr{L}{|\rho(t)\rangle}$, the vectorized density matrix $
|\rho\rangle=\sum_{m, n} \rho_{m, n}|m\rangle \otimes|n\rangle
$ where $\rho_{m, n}=\langle m|\rho| n\rangle$ is matrix element of the density matrix. The connected Liouvillian superoperator is represented as the following form
\begin{align}
\mathscr{L}= & -i\left(H \otimes I-I \otimes H^{\mathrm{T}}\right) \nonumber \\
	& +\sum_j\Gamma_{j}\left[2 \ell_{j} \otimes \ell_{j}^*-\ell_{j}^{\dagger} \ell_{j} \otimes I-I \otimes\left(\ell_{j}^{\dagger} \ell_{j}\right)^{\mathrm{T}}\right]. \label{Superoperator}
\end{align}
Where $I$ denotes the unit matrix in the Hilbert space of the system, and hence the dissipative dynamics and the steady state are fully determined by the spectrum of the Liouvillian superoperator, which can be obtained by directly diagonalizing \eqref{Superoperator}.

The spectrum properties of the Liouvillian superoperator \eqref{Superoperator} depend on the choice of operators $\ell_{j}$. In this work, we consider the following jump operator for the Eq. (\ref{Dj}), which is given as
\begin{eqnarray}\label{jumpoperator}
	\ell_{j}=(c^{\dagger}_{j}+e^{i\theta}c_{j+l}^{\dagger})(c_{j}-e^{i\theta}c_{j+l}).
\end{eqnarray}
The jump operator $\ell_{j}$acts on the  $j$-th and $j+l$-th sites synchronously. One can see the jump operator does not alter  particle number of the system but the relative phase between the site  $j$ and  the site $j+l$. For instance, it synchronizes the pair of sites from an in-phase  (out-of-phase) mode to an out-of-phase (in-phase) mode when the phase meets the condition $\theta=\pi$ ($\theta=\pi$). We will see that this property is indispensable for obtaining the intended ergodic or nonergodic stationary states.

Notably, it has been found that dephasing noise can spoil Anderson localization and enhances transport for a disordered system. It is generally acknowledged that dissipation tends to destroy the localization behavior and drives the system to a steady state without localization. However,  recently, Wang~\cite{wei2024coexistence} and coworkers applied the jump operator (\ref{jumpoperator}) to a 1D quasiperiodic system with MEs and discovered that the dissipation can drive the system into specific states, which may be extended or localized, regardless of the choice of initial state. This reveals that the dissipation can induce the transition between extended and localized states. In the next section, we investigate this type of dissipation in disorder-free systems with pseudo-MEs, which signifies the critical energy that distinguishes between ergodic and weakly ergodic, or weakly ergodic and nonergodic regions.

\begin{figure*}[t]
	\centering
	\includegraphics[width=0.65\textwidth]{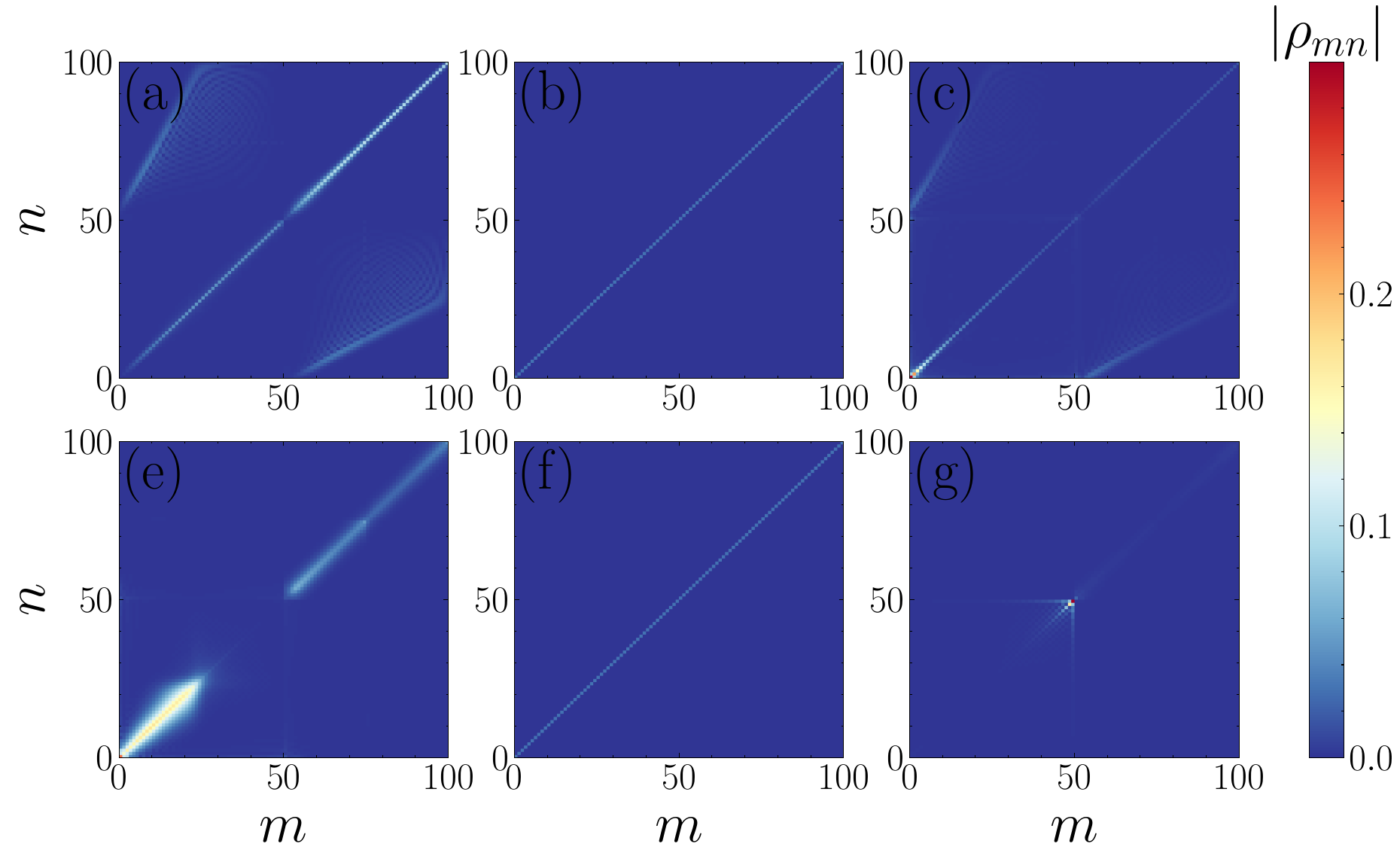}
    \includegraphics[width=0.31\textwidth]{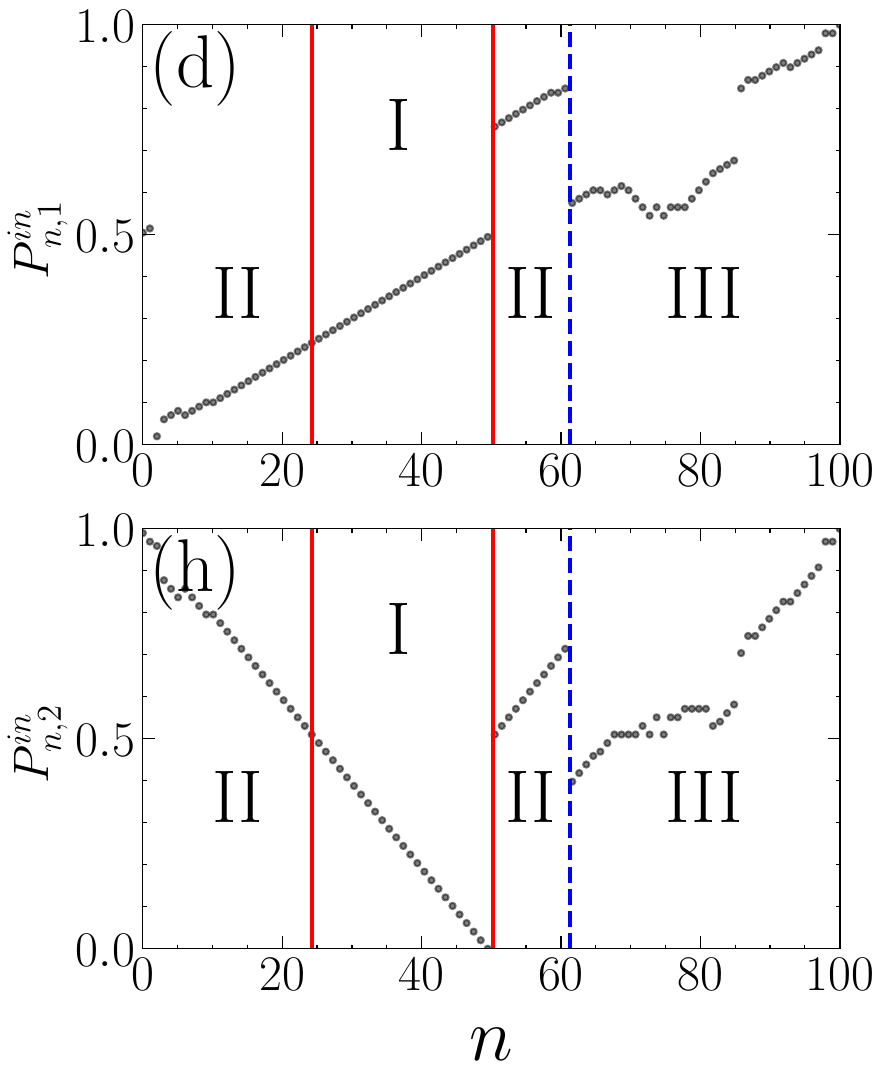}
	\caption{The density matrix distributions for steady states with the different dissipation phases (a) $\theta=0$, (b) $\theta=\pi/2$, and (c) $\theta=\pi$ in the eigenstates basis of the Hamiltonian (\ref{Hamiltonian}) with $\kappa=2$ and $l=1$. (d) The fraction of in-phase pairs $P_{n,1}^{\text{in}}$ for different eigenstates. The red solid lines mark the pseudo-MEs, separating eigenstates into weak ergodic (II), ergodic (I), and weak ergodic (II) regions as eigenvalues increase. The density matrix distributions for steady states with the different dissipation phases (e) $\theta=0$, (f) $\theta=\pi/2$, and (g) $\theta=\pi$ in the eigenstates basis of the Hamiltonian (\ref{Hamiltonian}) with $\kappa=2$ and $l=2$. (h) The fraction of in-phase pairs $P_{n,2}^{\text{in}}$ for different eigenstates. Here we take $L=100$, $\Gamma=1$, and $F_{max}/J=20$.}
	\label{fig4}
\end{figure*}

\begin{figure}[b!]
	\centering
	\includegraphics[width=0.48\textwidth]{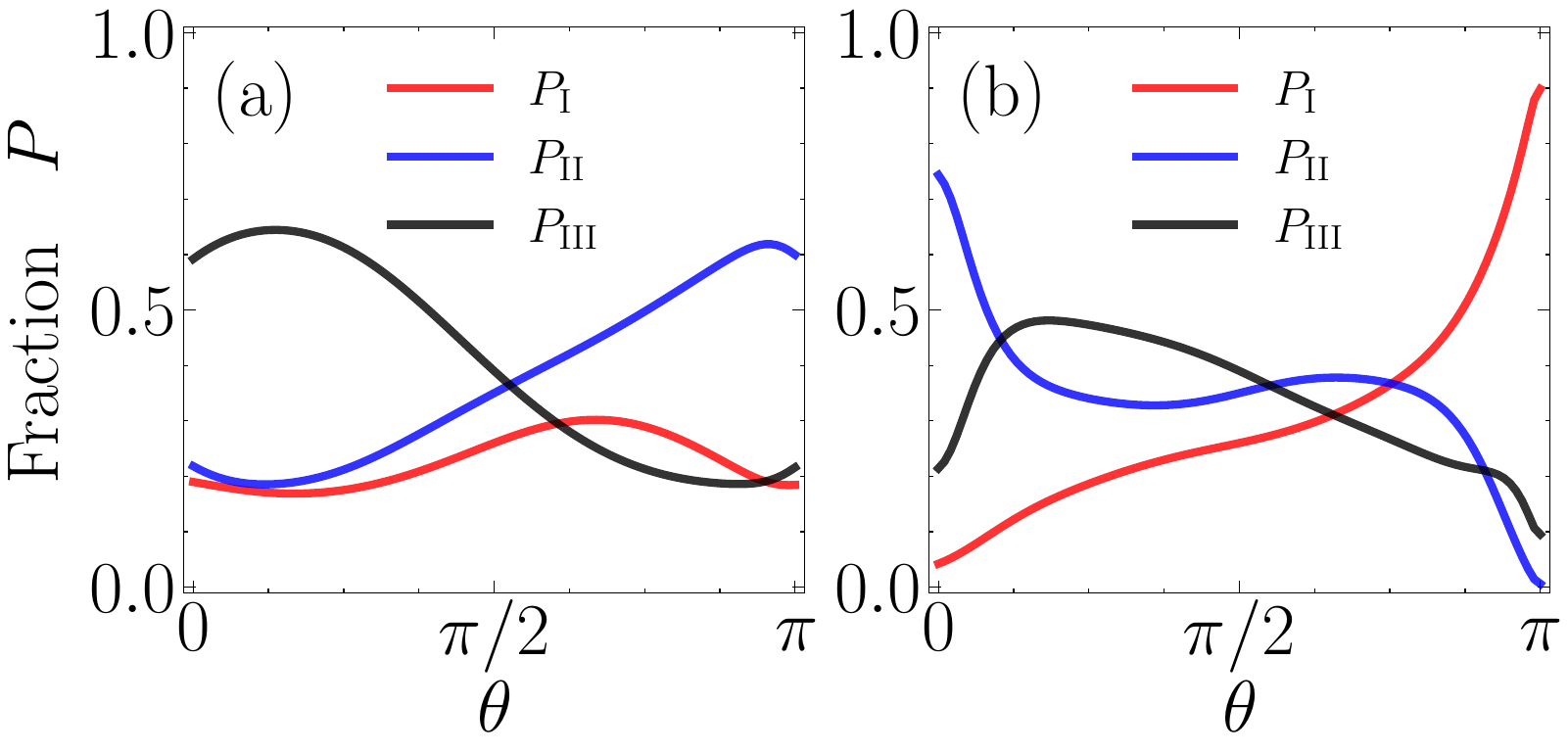}
	\caption{Fractions of ergodic eigenstates $P_{\rm{I}}$, weakly ergodic eigenstates $P_{\rm{II}}$ and nonergodic eigenstates $P_{\rm{III}}$ in corresponding steady states as the function of dissipation phase $\theta$ for the Hamiltonian (\ref{Hamiltonian}) with $\kappa=2$. (a) and (b) show the results for $l=1$ and $l=2$ case, respectively.}
	\label{fig5}
\end{figure}

\section{Numerical results and discussion}\label{sec: results}

\subsection{$\kappa=1$ case}
We first discuss the situation of jump operators in Eq. (\ref{jumpoperator}) takes $l=1$ and analyze the stationary state properties in the eigenstate representation of the Hamiltonian $H$, i.e. $\rho_{mn}=\langle \psi_m|\rho_{ss}|\psi_n\rangle$ with eigenstates  $|\psi_m\rangle$ and $|\psi_n\rangle$ of the Hamiltonian $H$. Figs.~\ref{fig2}(a-c) show the steady state transition from being mainly consisted of weakly ergodic states in the high energy regime to predominantly consists of  weakly ergodic states in the low energy regime as the phase varies from $\theta=0$ to $\theta=\pi$. Since the steady state is independent of the choice of the initial state, that implies if the initial state is chosen in the ergodic region located in the middle of the spectrum, the system ultimately relaxes to the steady state concentrating on weakly ergodic regions when the dissipation phase it set as $\theta=0$ or $\theta=\pi$.
	
To understand the origin of steady states composited of weakly ergodic states, we analyze relative phases between a pair of neighboring sites. Specifically, for the $n$-th eigenstate $|\psi_n\rangle=\sum_{j}\psi_{n,j}c_{j}^{\dagger}|0\rangle$, we compute the phase difference $\Delta\theta^n_{j,l}$ between the $j$-th and the $(j+l)$-th site as $\Delta\theta^n_{j,l}=\arg(\psi_{n,j})-\arg(\psi_{n,j+l})$. If the phase difference is zero, i.e. $\Delta\theta^n_{j,l}=0$, we call it in phase. Hence, one can calculate the in phase site pairs number $N_{n,l}^{\text{in}}$, and its proportion $P_{n,l}^{\text{in}}=N_{n,l}^{\text{in}}/N_t$ with $N_t=L-l$ being total number of pairs. Fig.~\ref{fig2}(d) shows that the eigenstate with lower (higher) energy prefers to gain smaller (larger) $P_{n,1}^{\text{in}}$, which is why the steady state mainly concentrates on weakly ergodic eigenstates with low-energy (high-energy) for $\theta=\pi~(\theta=0)$.

We further explore the situation of jump operators in Eq. (\ref{jumpoperator}) with $l=2$. We first compute the proportion of in phase site pairs $P_{n,2}^{\text{in}}$. We find that it displays a $V$-shaped pattern as shown in~[Fig.~\ref{fig2}(h)]. Weakly ergodic states on both sides of the spectrum have more in phase pairs, while the ergodic states in middle of the spectrum tend to obtain more out of phase pairs. Therefore, by choosing the dissipation phase appropriately, one can control whether the steady state of the system is composed of weakly ergodic or ergodic eigenstates, as plotted in Figs.~\ref{fig2}(e-g). When the dissipation phase is set $\theta=0$ as shown in the Fig.~\ref{fig2}(e)], the system relaxes to a steady state mainly composed of states associated with in phase site pairs, and thus concentrating on the weakly ergodic eigenstates in both lower energy and higher energy regions. By contrast, when $\theta=\pi$ as shown in Fig.~\ref{fig2}(g), the system is expected to reach another steady state predominantly consisting of those states with out-of-phase pairs and prefers ergodic eigenstates in the middle spectrum region. Next we choose the dissipation phase as $\theta=\pi/2$. As shown in the Fig.~\ref{fig2}(f) and the Fig.~\ref{fig2}(b), we see that the system reaches to the infinite high temperature state or maximally mixed state  $(\rho_{ss})_{mn}=\delta_{mn}/L$. This is because dissipation operators are Hermitian, and the steady state density matrix must be proportional to the identity matrix which commutes with dissipation operators. Furthermore, based on diagonal elements of the density matrix, we can determine proportions of ergodic and weakly ergodic eigenstates in the steady state for arbitrary $\theta$, i.e., $P_{\rm{I}}=\sum_{j}\rho_{jj}$ ($P_{\rm{II}}=\sum_k\rho_{kk}$), where $j (k)$ denotes the index of ergodic (weakly ergodic) eigenstates $|\psi_j\rangle$ ($|\psi_k\rangle$). However, when $\theta$ is tuned from $0$ to $\pi$ for the case of $l=2$, the steady state exhibits a transition from the state dominated by ergodic eigenstates to the state dominated by weakly ergodic eigenstates as shown in the Fig.~\ref{fig3}(b). This shows that dissipation can be used to manipulate the ergodic-weakly ergodic transition.

\subsection{$\kappa=2$ case}
In this subsection we investigate the Hamiltonian (\ref{Hamiltonian}) with $\kappa=2$ case, which has two types of pseudo-MEs. When we consider the parameter $F_{max}/J=20$, the region coexists in ergodic, weakly ergodic, and nonergodic eigenstates. The eigenstates go through II$\rightarrow$I$\rightarrow$II$\rightarrow$III, as shown in Fig.~\ref{fig1}(b). We follow the process that is similar to the $\kappa=1$ case. We first investigate the jump operators in Eq. (\ref{jumpoperator}) for $l=1$, then analyzing properties of the stationary state. In Figs.~\ref{fig4}(a-c),  we plot the steady state transition from mainly consisting of high-energy nonergodic states to predominantly consisting of low-energy weakly ergodic states as the dissipation phase varies from $\theta=0$ to $\theta=\pi$. Since the steady state is independent of the choice of the initial state, that implies if the initial state is chosen in the ergodic region located in the I region, the system ultimately relaxes to the steady state concentrating on high-energy nonergodic or low-energy weakly ergodic regions when the dissipation phase it set as $\theta=0$ or $\theta=\pi$.

Therefore, by applying this type of dissipation, the ergodic state can be driven to a steady state consisting mainly of weakly ergodic or non-ergodic states. In the same way as discussed for $\kappa=1$ case, we can understand composition of steady states by examining relative phases between a pair of neighboring sites. Fig.~\ref{fig4}(d) plots the proportion of in-phase pairs $P_{n,1}^{\text{in}}$ and shows that 
the eigenstate with  higher (lower) energy prefers to gain smaller (larger) $P_{n,1}^{\text{in}}$, which is why the steady state mainly concentrates on high-energy nonergodic (low-energy weakly ergodic) eigenstates for $\theta=0~(\theta=\pi)$.



We then study the case of $l=2$. As shown in Fig.~\ref{fig4}(h), we compute the proportion of in-phase site pairs $P_{n,2}^{\text{in}}$, and find that it exhibits a discontinuous $V$-shaped pattern. The weakly ergodic and nonergodic states on both sides of the energy spectrum have more in-phase site pairs, while the ergodic states in the middle of the pseudo-MEs (\ref{k2}) tend to gain more out-of-phase pairs. 
Therefore, by choosing the dissipation phase appropriately, one can  control whether the steady state of the system is composed of weakly ergodic or ergodic eigenstates, as plotted in Figs.~\ref{fig4}(e-g). When the dissipation phase is set $\theta=0$ as shown in the Fig.~\ref{fig4}(e)], the system relaxes to a steady state mainly composed of states associated with in-phase site pairs, and hence primarily concentrating on the weakly ergodic eigenstates in the lower-energy region. By contrast, when $\theta=\pi$ as shown in Fig.~\ref{fig4}(g), the system is expected to reach another steady state primarily consisting of those states with out-of-phase pairs and prefers ergodic eigenstates in the middle spectrum region. For the case of dissipation phase $\theta=\pi/2$, the system ultimately reaches to the infinite high temperature state or maximally mixed state  $(\rho_{ss})_{mn}=\delta_{mn}/L$ as shown in the Fig.~\ref{fig4}(b) and the Fig.~\ref{fig4}(f).

Therefore, we can manipulate whether the steady state of the system is composed of weakly ergodic or ergodic eigenstates, as plotted in Figs.~\ref{fig4}(e-g). When the dissipation phase is set $\theta=0$ as shown in the Fig.~\ref{fig4}(e)], the system relaxes to a steady state mainly composed of those states with in-phase pairs, and thus concentrating on the weakly ergodic eigenstates in  the lower-energy region. By contrast, for the case of $\theta=\pi$ as shown in Fig.~\ref{fig4}(g), the system reaches another steady state predominantly consisting of those states with out-of-phase pairs and prefers ergodic eigenstates in the middle spectrum region. 

For arbitrary $\theta$, we can determine the fractions of distinct states in steady states, i.e., $P_{\rm{I}}=\sum_{i}\rho_{ii}$ ($P_{\rm{II}}=\sum_j\rho_{jj}$,$P_{\rm{III}}=\sum_k\rho_{kk}$), where $i (j,k)$ is the index of ergodic (weakly ergodic, nonergodic) eigenstates $|\psi_i\rangle$ ($|\psi_j\rangle$,$|\psi_k\rangle$).

However, when $\theta$ varies from $0$ to $\pi$ for the case of $l=2$, the steady state exhibits a transition from the state dominated by nonergodic eigenstates to the state dominated by weakly ergodic eigenstates as shown in the Fig.~\ref{fig5}(a). This shows that dissipation can be used to manipulate the weakly ergodic-nonergodic transition. As shown in Fig.~\ref{fig5}(b), we also see that dissipation can be used to control the weakly ergodic to ergodic transition.

In the context of the aforementioned findings, we have exhibited that an ergodic (weakly ergodic) state can be steered towards a stable state, predominantly constituted of weakly ergodic (ergodic or nonergodic) states, through the manipulation of dissipation. By introducing and controlling a relative dissipation phase, the system undergoes a profound transition while the Hamiltonian parameters remain unchanged. Therefore, the dissipation provides a mechanism to manipulate transitions among different states, specifically between ergodic and weakly ergodic states, or between weakly ergodic and nonergodic states.
%
%

\section{Conclusion}\label{sec: conclusion}
We have explored the impact of dissipation on the one-dimensional finite-height mosaic Wannier-Stark model possessing pseudo-MEs that distinguish between ergodic and weakly ergodic or weakly ergodic and nonergodic states. By studying distributions of the density matrix of steady states and the proportion of in-phase pairs, we found that dissipation is able to drive this system into specific states mainly occupied by ergodic, weakly ergodic, or nonergodic states without depending on the initial states. Hence, one can utilize dissipation to induce transitions between ergodic and nonergodic states in disorder-free systems with pseudo-MEs, thereby manipulating particle transport behaviors. We expect that the manipulation of ergodic-nonergodic transitions can be applied to those interacting systems exhibiting MEs. Our findings also prompt several exciting issues. For instance, how does this dissipation affect a Stark many-body system with MEs? Can this kind of dissipation manipulate the transition between ergodic states and Stark many-body localized states? Further, the dissipation operator in Eq. (\ref{jumpoperator}) employs the phase characteristics of distinct states in the spectrum to extract specific states as a steady state. Our results seem to provide a new perspective for exploring experimentally feasible dissipation operators to select different quantum states.

\section*{Acknowledgments}
We thank Yucheng Wang for useful discussions. This work is supported by the China Postdoctoral Science Foundation (No.~2023M743267) and the National Natural Science Foundation of China (Grant No.~12304290 and No.~12204432). LP also acknowledges support from the Fundamental Research Funds for the Central Universities. 

\bibliography{Localization}
\end{document}